\documentclass[
footinbib,
a4paper,
aps,
prx,
superscriptaddress,
reprint,
twocolumn,
preprintnumbers,
nobalancelastpage,
floatfix,
longbibliography,
10pt]{revtex4-2}

\usepackage{xcolor}
\usepackage{graphicx}
\usepackage{dcolumn}

\usepackage{subfigure}
\usepackage{float}

\usepackage[utf8]{inputenc}

\usepackage{amsmath, amsthm, amsfonts, amssymb,amstext}
\usepackage{mathtools}
\usepackage{bm}

\usepackage{braket}
\usepackage{siunitx}

\definecolor{cmblue}{rgb}{0.12156862745098039, 0.4666666666666667, 0.7058823529411765}
\definecolor{myblue}{rgb}{0.2,0.2,0.8}
\usepackage[
colorlinks=true,
citecolor=myblue,
linkcolor=cmblue,
urlcolor=myblue,
pdftitle={Characterizing physical and logical errors in a transversal CNOT via cycle error reconstruction}]{hyperref}
\usepackage{cleveref}
\usepackage{verbatim}
\newcommand{\cnot}{\textsc{cnot} }
\newcommand{\cnots}{\textsc{cnot}s }

\usepackage{silence}
\begin{document}

\title{Characterizing physical and logical errors in a transversal CNOT\\ via cycle error reconstruction}

\author{Nicholas Fazio}
    \altaffiliation{These authors contributed equally}
\affiliation{Centre for Engineered Quantum Systems, School of Physics, The University of Sydney, Sydney, NSW 2006, Australia}

\author{Robert Freund}
    \altaffiliation{These authors contributed equally}
\affiliation{Universit{\"a}t Innsbruck, Institut f{\"u}r Experimentalphysik, Innsbruck, Austria}

\author{Debankan Sannamoth}
\affiliation{Department of Applied Mathematics, University of Waterloo, Waterloo, Ontario N2L 3G1, Canada}
\affiliation{Institute for Quantum Computing, University of Waterloo, Waterloo, Ontario N2L 3G1, Canada}

\author{Alex Steiner}
\affiliation{Universit{\"a}t Innsbruck, Institut f{\"u}r Experimentalphysik, Innsbruck, Austria}

\author{Christian D. Marciniak}
\affiliation{Universit{\"a}t Innsbruck, Institut f{\"u}r Experimentalphysik, Innsbruck, Austria}

\author{Manuel Rispler}
\affiliation{Institute for Quantum Information, RWTH Aachen University, Aachen, Germany}
\affiliation{Institute for Theoretical Nanoelectronics (PGI-2), Forschungszentrum J\"{u}lich, J\"{u}lich, Germany}

\author{Robin Harper}
\affiliation{Centre for Engineered Quantum Systems, School of Physics, The University of Sydney, Sydney, NSW 2006, Australia}

\author{Thomas Monz}
    \altaffiliation[Also at: ]{Alpine Quantum Technologies GmbH, Innsbruck, Austria}
\affiliation{Universit{\"a}t Innsbruck, Institut f{\"u}r Experimentalphysik, Innsbruck, Austria}

\author{Joseph Emerson}
\affiliation{Department of Applied Mathematics, University of Waterloo, Waterloo, Ontario N2L 3G1, Canada}
\affiliation{Institute for Quantum Computing, University of Waterloo, Waterloo, Ontario N2L 3G1, Canada}

\author{Stephen D. Bartlett}
\affiliation{Centre for Engineered Quantum Systems, School of Physics, The University of Sydney, Sydney, NSW 2006, Australia}

\date{\today}


\begin{abstract}
The development of prototype quantum information processors has progressed to a stage where small instances of logical qubit systems perform better than the best of their physical constituents. Advancing towards fault-tolerant quantum computing will require an understanding of the underlying error mechanisms in logical primitives as they relate to the performance of quantum error correction. 
In this work we demonstrate the novel capability to characterize the physical error properties relevant to fault-tolerant operations via \textit{cycle error reconstruction}. We illustrate this diagnostic capability for a transversal \textsc{cnot}, a prototypical component of quantum logical operations, in a 16-qubit register of a trapped-ion quantum computer. 
Our error characterization technique offers three key capabilities:
(i)~identifying context-dependent physical layer errors, enabling their mitigation; 
(ii)~contextualizing component gates in the environment of logical operators, validating the performance differences in terms of characterized component-level physics, and
(iii)~providing a scalable method for predicting quantum error correction performance using pertinent error terms, differentiating correctable versus uncorrectable physical layer errors.
The methods with which our results are obtained have scalable resource requirements that can be extended with moderate overhead to capture overall logical performance in increasingly large and complex systems.
\end{abstract}

\maketitle

\section{Introduction}
\label{sec:Introduction}
It is widely believed that a quantum computer will need to employ quantum error correction (QEC) in order to operate at sufficient scale to run algorithms of practical interest~\cite{beverland2022,dalzell2025}. 
The goal of realizing a fault-tolerant, digital quantum computer has recently seen great experimental progress in different hardware platforms employing QEC from the first universal fault-tolerant gate sets~\cite{postler2022}, encoded logical circuits~\cite{bluvstein2024logical}, beyond-break-even magic state generation~\cite{gupta2024encoding}, to operation of codes where logical state preparation and quantum memories perform better than the best unencoded constituent qubits~\cite{reichardt2024demonstration,acharya2024quantum}.

These early demonstrations of QEC have motivated the need for new characterization tools, to verify the functionality of QEC and to validate the fault-tolerant operation of logical components. A significant challenge for any such tool is in how to characterize logical devices at a scale reaching tens or hundreds of physical qubits. 
Early quantum characterization, verification, and validation (QCVV) techniques such as randomized benchmarking~\cite{emerson2005scalable,magesan2011scalable,magesan2012characterizing}, process tomography~\cite{Weinstein2004QuantumProcess} and gate set tomography~\cite{BlumeKohout2013Robust,Nielsen2021GateSet} have been applied primarily on the smallest building blocks of quantum circuits or provide highly condensed information for larger blocks. Recently, several techniques have emerged that are based on the idea of characterizing blocks of quantum operations such as \textit{layers} or \textit{cycles} of parallel operations across many qubits to extract representative error descriptions that account for crosstalk and idling errors~\cite{erhard2019characterizing,carignandugas2023error,hockings2024scalable}. These techniques deliver a flexible error characterization framework based on a resource cost versus information gain tradeoff, enabling QCVV for intermediate- to large-scale quantum systems that includes both component characterization and full-system characterization methods suitable for current devices. 

Characterizing logical operations and their errors has been studied in the literature, for example by promoting physical layer techniques to encoded variants~\cite{wagner2023learning}.
The transition from physical layers to logical layers is not trivial since logical qubits and the QEC machinery needed for them can generate violations of common error-model assumptions for current QCVV techniques~\cite{ceasura2022non}.
Encoded qubits possess features that are not dominant or are altogether absent on the physical layer such as repeated stabilizer readout via mid-circuit measurements, and future operations conditioned on the outcomes of earlier measurements.
While recent proposals have made progress on benchmarking these features~\cite{govia2023randomized,zhang2024generalized,Hines2025,harper2025characterising}, to what extent useful information can be gained about logical performance across physical layers remains a central challenge for QCVV of error-corrected systems.

In this work we apply \textit{cycle error reconstruction} (CER)~\cite{carignandugas2023error}, a method for scalably reconstructing error characteristics, to characterize a logical primitive. Specifically, we demonstrate this characterization method for a transversal \cnot on a pair of logical qubits realized with the 7-qubit Steane code~\cite{steane1996multiple} implemented on a trapped-ion quantum computer~\cite{pogorelov2021,postler2022,postler2024demonstration}.
We theoretically develop and experimentally demonstrate three principal applications of this new `error characterization for QEC' capability: 
\begin{enumerate}
    \item[(i)] We demonstrate the effectiveness of CER to obtain detailed and statistically precise information about the error characteristics for quantum gate operations relevant to the successful execution of fault-tolerant logic.
    We optimize the resources required under an adjustable tradeoff between resource cost and information gain so that we can acquire accurate multi-qubit correlated Pauli error rates across a large array of qubits on a short time-frame for our experimental platform.
    \item[(ii)] With the data obtained from CER, we validate detailed error models for describing multi-physical-qubit systems in the context of logical operations.
    We demonstrate how the overall noise characteristics of the transversal \cnot are well understood in terms of the noise characteristics obtained via CER from individual (physical) \cnots together with well-characterized idling error rates.
    \item[(iii)] We demonstrate the predictive capacity of the CER approach by bounding the logical fidelity based on the learned errors.
    We extend the graphical noise model framework based on the Gibbs random field framework~\cite{harper2020efficient,harper2023learning} to account for the action of the transversal CNOT, and show that low Hamming weight CER data can be used to construct a global error model for the physical process.
    With our CER data, we use our global error model to predict and classify correctable versus uncorrectable error contributions of the transversal \cnot under quantum error correction. 
\end{enumerate}
In this study we focus on illustrating these capabilities in the context of a layer of transversal gate operations performed in a 16-qubit array of trapped ions, expecting that this approach will be applicable to broader contexts by adapting the global error models as appropriate to the specific logical gate and fault-tolerant architecture.

The manuscript is structured as follows: We begin in Sec.~\ref{sec:CER} with an overview of CER, focusing on the practical aspects -- including required resources -- involved with executing a suite of experiments.
We briefly outline the experimental platform in Sec.~\ref{sec:Experimental}. The main results of this work are presented in Sec.~\ref{sec:Results}: We apply CER to both individual physical \cnots as well as the transversal \textsc{cnot}, providing detailed error information of both the individual component gates and the overall transversal operation. We use this CER data to validate our error model, demonstrating that component-level characterization together with simple device physics can be used to predict system-level characteristics. We also show how CER data together with a Gibbs random field model can be used to make predictions about logical-level performance~\cite{harper2020efficient,harper2023learning}. This work concludes in Sec.~\ref{sec:Outlook} with future directions.


\section{Cycle Error Reconstruction}
\label{sec:CER}

We begin with an exposition of \textit{cycle error reconstruction}~\cite{carignandugas2023error}. We will use this technique to study the noise characteristics of a transversal \cnot\cite{east09}, as well as the 7 individual physical \cnots that constitute this gate.

Earlier benchmarking methods employed sequences of random gates, which effectively twirl the error by averaging it over a unitary 2-design~\cite{dankert2009exact}, to yield a single-number fidelity benchmark that is robust to state preparation and measurement errors (SPAM) at low resource cost~\cite{emerson2005scalable}.
Clifford randomized benchmarking~\cite{magesan2011scalable,magesan2012characterizing} is arguably the most well-known and common instance of this, where the $n$-qubit Clifford group is the relevant 2-design.
In a departure from this, cycle error reconstruction (CER) builds on the randomized, iterated circuit structure developed in \textit{cycle benchmarking}~\cite{erhard2019characterizing} and eigenvalue estimation techniques~\cite{Emerson2007,flammia2020efficient} to characterize fine-grained information about the error model affecting target cycles of interest.
These methods leverage the generalized twirling and compilation strategy of randomized compiling
~\cite{wallman2016noise}, which in our usage only requires randomizing over products of single-qubit operators rather than $n$-qubit Clifford operators.
The CER method generates multiple decay parameters and leverages these parameters to estimate large sets of Pauli error probabilities for the cycle of interest, which provides enhanced predictive and diagnostic utility.
For comparison, related methods such as \textit{averaged circuit eigenvalue sampling} and \textit{layer benchmarking} have been proposed to characterize cycle errors but are restricted to characterizing Clifford operations and the latter only produces a summative fidelity over cycle errors~\cite{mckay2023benchmarking,hockings2024scalable}.


\subsection{Setting and conventions} 

Let us first define terminology and objects that we will require in the following sections:
The eponymous \textit{cycle} of CER refers to a sequence of operations in a quantum circuit that occur `in parallel', that is, within one timestep determined by the operational capabilities of a device. 
These operations may be grouped together based on their ability to be parallelized in time on a device, or alternatively, we might require that each gate is supported on disjoint sets of qubits even though in an actual experiment each operation would need to be implemented serially. 
We divide these cycles into two types: `easy cycles' and `hard cycles'. 
Easy cycles are composed of well-characterized operations with relatively high fidelity for a given device, such as single-qubit rotations. Hard cycles will contain operations that are comparatively noisy, such as entangling gates or mid-circuit measurements, which we seek to characterize. Any circuit can be written as an alternating sequence of easy and hard cycles.

\begin{figure}[t]
    \centering
    \includegraphics[width=8.5cm]{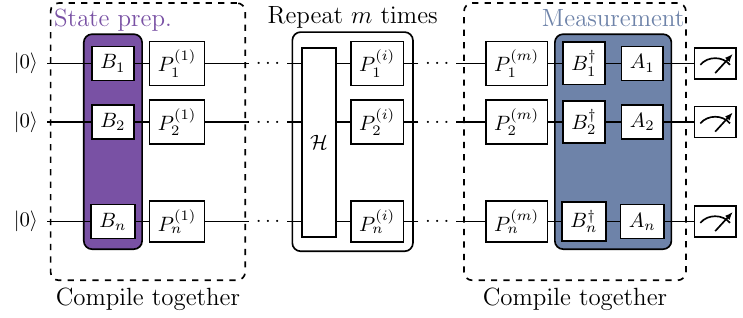}
    \caption{Cycle benchmarking circuit design \cite{erhard2019characterizing} leveraged to learn individual Pauli eigenvalues to relative precision. Illustrated in this example is the simpler form that is sufficient for any Clifford hard cycle $\mathcal{H}$ where easy cycles are Pauli randomized cycles $P$~\cite{carignandugas2023error}. The easy cycle $B$ includes local rotations to prepare Pauli product states and define the measurement basis. The randomized Pauli cycle $A$ eliminates bias from measurement errors.}
    \label{fig:pauli-eigenvalue-learning}
\end{figure}

In \textit{cycle error reconstruction} we aim to characterize the errors on these noisier hard cycles. 
The hard cycle of interest is applied to a set of initial states and measured over different bases for a variable number of cycles.
These hard cycles $\mathcal{H}$ are interleaved between easy gates $P$, where the easy gates are randomized in order to enforce a Pauli twirl, as schematically depicted in Fig.~\ref{fig:pauli-eigenvalue-learning}. 
This division between easy and hard cycles ensures that any systematic error associated with the interleaved easy cycles will be small relative to the error on the hard cycle of interest.

In order to learn the profile of the noise introduced by the hard cycle we need to learn various Pauli eigenvalues, whose definition and estimation are outlined in Sec.~\ref{sec:pauli-eigenvalue}. 
Given estimates for the Pauli eigenvalues we are then able to infer quantities of interest, namely, the Pauli error probabilities of the hard cycle.

For clarity we will provide some basic definitions and conventions: We define the single-qubit Pauli group $\mathbb{P}$ as the set of single-qubit Pauli operators with equivalence up to global phase \mbox{$\mathbb{P} = \mathbb{P}_1  \coloneqq \{I,  Z, X, Y\}$}. 
We likewise define the $n$-qubit abelian Pauli group~$\mathbb{P}_n$, whose elements are the tensor product of single-qubit Pauli operators with equivalence up to global phase. We may write elements of~$\mathbb{P}_n$ in terms of products of elements in~$\mathbb{P}_1$, indexed by the qubit they act on, e.g. $P = Z_iY_jY_k\cdots$ has $Z$ act on qubit $i$, $Y$ act on qubit $j$ and so on.

We will also consider the restriction of elements in~$\mathbb{P}_n$ to arbitrary subsets of qubits.
The restriction of Pauli~$P$ to the qubits indexed by elements of set~$S$ is denoted $P_S$. For example, $[Z_1X_2Y_4]_{\{1\}} = Z_1$, $[Z_1X_2Y_4]_{\{2,3\}} = X_2I_3$, or $[Z_1X_2Y_4]_{\{4,1\}} = Y_4Z_1$.
Note that the ordering of indices in the set is used here to typographically order the restricted Pauli.
Similarly, we denote by $\mathbb{P}_S$ the subgroup of $\mathbb{P}_n$ that is supported on only the qubits of~$S$.
It will also be useful in this context, to define the map~$\omega$ between arbitrary Pauli operators~$P$ and~$Q$, defined as
\begin{equation} \omega(P,Q) = \begin{cases} 
      1, & \text{$P$ and $Q$ anticommute} \\
      0, & \text{$P$ and $Q$ commute}.
   \end{cases} \label{eq:omega}
\end{equation}

The $n$-qubit Pauli operators form a Hermitian basis for Hilbert space, so any quantum channel $\mathcal{G}$ can be expressed in terms of the Pauli basis as
\[
    \mathcal{G}(\rho) = \sum_{P_i,P_j \in \mathbb{P}_n}\chi_{ij}P_i\rho P_j^\dag,
\]
where $\chi_{ij}$ is the so-called \textit{process} or $\chi$~\textit{matrix} of coefficients. 
The $\chi$-matrix generally has $\sim16^n$ non-trivial parameters, presenting one of the central challenges of QCVV: to characterize quantum channels given that they have exponentially many parameters. 
A special case of $\mathcal{G}(\rho)$ is the stochastic Pauli channel, which can be expressed in the form
\begin{equation}
    \mathcal{P}(\rho) = \sum_{P_i \in \mathbb{P}_n} p_iP_i\rho P^\dag_i,
    \label{eq:pauli-channel}
\end{equation}
where $ p_i = \chi_{ii}$ is the classical probability of some Pauli operator $P_i$ occurring in the quantum channel. Evidently, the number of non-trivial parameters in $\mathcal{P}$ is drastically smaller than in $\mathcal{G}$; at most $4^n-1$.
These smaller operators are often more practical to study.

Many common sources of noise fall outside the class of stochastic Pauli noise models. However, an arbitrary channel can be tailored into a stochastic Pauli channel by averaging or `twirling' over the Pauli group such that
\[
    \mathcal{G}^{\mathbb{P}_n}(\rho) = \frac{1}{|\mathbb{P}_n|}\sum_{P_i \in \mathbb{P}_n} P_i \mathcal{G}(P_i^\dag \rho P_i)P_i^\dag= \sum_{P_i \in \mathbb{P}_n} \chi_{ii}P_i\rho P^\dag_i,
\]
where $|\mathbb{P}_n|$ is the cardinality of $\mathbb{P}_n$. By averaging over the Pauli group in this manner the off-diagonal $i\neq j$ terms of the channel are eliminated so that the average statistics are described by a stochastic Pauli channel $\mathcal{G}^{\mathbb{P}_n}$.

Channels of interest are generally hard cycles that are not the identity or idling operation. 
Typically we have some ideal channel~$\mathcal{H}$ that is implemented imperfectly as $\tilde{\mathcal{H}} = \mathcal{E} \circ \mathcal{H}$ where~$\mathcal{E}$ is the noise channel we are interested in quantifying. 
We are thus not always able to insert random Pauli operators either side of the noise channel $\mathcal{E}$, but we may still interleave operations either side of~$\tilde{\mathcal{H}}$. 
We can effectively twirl $\mathcal{E}$ if we instead insert a Pauli operator $P_i^c = \mathcal{H}^\dag (P_i)$ before $\tilde{\mathcal{H}}$ for each random~$P_i$ inserted after~$\tilde{\mathcal{H}}$.

The procedure of inserting a random Pauli on one side of a gate, a corresponding conjugate operator on the other side of the gate, and then compiling these operations together into a single layer is known as \textit{randomized compiling}~\cite{wallman2016noise}. 
When the hard cycles $\mathcal{H}$ are composed of Clifford gates, $P_i^c$ will be a Pauli operator and compiles together with other Pauli operators to give a single Pauli cycle between each hard cycle. 
The hard cycles we study in this work -- single and transversal \cnots -- satisfy this criterion. 
Through usage of randomized compiling we will assume all average-noise is described by Pauli channels unless otherwise stated. 
We also assume throughout that the noise channel introduced by~$\tilde{\mathcal{H}}$ is time-independent, i.e., we do not account for drift.


\subsection{Pauli Eigenvalue Learning} \label{sec:pauli-eigenvalue}

The \textit{Pauli eigenvalue} $\lambda_{P_i}$ of a Pauli channel $\mathcal{P}$ is a parameter that describes the change in the Pauli $P_i$ expectation value resulting from an arbitrary state $\rho$ passing through this noise channel, that is,
\begin{equation}
    \langle P_i\rangle_{\mathcal{P}(\rho)} = \text{tr}(P_i\mathcal{P}(\rho)) = \lambda_{P_i}\text{tr}(P_i \rho) =\lambda_{P_i}\langle P_i\rangle_{\rho}.
    \label{eq:eigval-def}
\end{equation}
The parameters $\lambda_{P_i}$ correspond to the matrix eigenvalues of a Pauli noise channel in the Liouville superoperator representation, which motivates the naming convention. 

If we express $\mathcal{P}$ in terms of Kraus operators, as in Eq.~\eqref{eq:pauli-channel}, it is straightforward to show that
\begin{equation}
    \lambda_{P_i} = \sum_{P_j\in \mathbb{P}_n} (-1)^{\omega(P_i,P_j)} p_j,
    \label{eq:pauli-eigenvalue}
\end{equation}
which is a discrete Fourier transform of the Pauli error probabilities $p_j$~\cite{zhang2024generalized, merkel2021randomized, helsen2019reptheory}: a linear transformation called the Walsh-Hadamard transform.
(In other works $\lambda_{P_i}$ has been called the Pauli fidelity~\cite{chen2023learnability, carignandugas2023error, zhang2024generalized}, but it would be more appropriately called a quasi-fidelity since~$\lambda_{P_i}$ can be negative and fidelities are typically non-negative quantities.)

The eigenvalues are determined in a manner that is robust to SPAM by the same noise amplification common to other randomized methods, that is repeated application $\mathcal{P}^m(\rho)$ of the hard cycle to some initial state~$\rho$ for different values of~$m$. The eigenvalues are then obtained by fitting the decay~$\lambda_{P_i}^m$ of the Pauli expectation value~$\langle P_i\rangle$.

\begin{figure}[ht!]
    \centering
    \includegraphics[width=8.5cm]{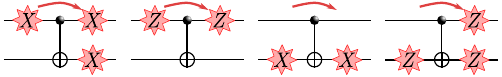}
    \caption{Circuit representation of how a single Pauli error (red stars) propagates through a single \cnot thereby changing its profile. Since errors occur probabilistically it is impossible to distinguish one type of error $P_i$ before the gate from the corresponding error $\mathcal{H}(P_i)$ after the gate.}
    \label{fig:Orbits}
\end{figure}

Hard cycle operations will generally act non-trivially on Pauli operators, meaning that each source of noise is also affected by the hard cycles~$\mathcal{H}$ that follow it. 
A Pauli error $P_i$ before a hard cycle $\mathcal{H}$ would be entirely indistinguishable from an error $\mathcal{H}(P_i)$ occurring after the hard cycle, as shown in Fig.~\ref{fig:Orbits}.

The hard cycles we use in this work are elements of the Clifford group, meaning that $\mathcal{H}(P_i) = P_j \in\mathbb{P}_n$ and that $\mathcal{H}^r=\mathcal{I}$ is the identity channel for finite $r$ since the Clifford group is finite.
Hence there is a finite set of distinct Pauli eigenvalues that affect the decay of a given Pauli expectation value, rather than just one.
To capture this relation, we define the \textit{orbit} $P^{\circlearrowright\mathcal{H}}$ under the action of~$\mathcal{H}$ for each Pauli $P$, as follows~\cite{carignandugas2023error}:
\begin{equation}
    P^{\circlearrowright\mathcal{H}} \coloneqq  \{\mathcal{H}^k(P) \mid k\in\mathbb{Z}_r\}.
    \label{eq:orbit}
\end{equation}

Unless~$P_i$ is invariant under the action of~$\mathcal{H}$, that is \mbox{$\mathcal{H}(P_i)=P_i$}, the observed decay parameter associated with~$\langle P_i\rangle$ will not simply be~$\lambda_{P_i}$ of~$\mathcal{P}$. 
Instead the action of the hard cycle means we observe the ensemble-averaged Pauli \textit{orbital} eigenvalue; the geometric mean of all the Pauli eigenvalues in the orbit of~$P$. 
This composite decay parameter is given by
\begin{equation}
    \lambda_{P^{\circlearrowright\mathcal{H}}} = \left(\prod_{P_i\in P^{\circlearrowright\mathcal{H}}}\lambda_{P_i} \right)^{1/|P^{\circlearrowright\mathcal{H}}|}.
    \label{eq:orbit-eigval}
\end{equation}  
Accordingly, all Pauli expectation values that share such an orbit will have the same decay rate, expressing a degeneracy in the learnable decay rates.

There are methods that can be used to further resolve the constituent Pauli eigenvalues out of the orbital eigenvalues~\cite{carignandugas2023error, chen2023learnability}.
However, there are fundamental limits on the learnability of Pauli noise~\cite{chen2023learnability} that demonstrate we cannot do this for all Pauli eigenvalues while also achieving \textit{relative precision}~\cite{harp19, carignandugas2023error, flammia2020efficient}, meaning that the number of shots required to reduce the uncertainty in the estimated error rates will grow as those error rates get smaller.
Here we will adopt the inherent cycle averaging as a feature -- rather than a problem to address -- allowing us to reduce the required number of independent parameters to learn while also maintaining relative precision. 
Relative precision in the estimation of a decay parameter~$\lambda$ is achieved by letting the largest number of cycle repetitions~$m$ satisfy $m_L \sim \frac{1}{1-|\lambda|}$. 
In essence this means repeating the noise channel an appropriate number of times so that the aggregate decay is sufficiently large. 
Initial fidelity estimates can be used to determine the relevant order of magnitude for~$m_L$.
Thereafter, ongoing eigenvalue estimates can be used to refine the largest value $m_L$.
It is important to note that relative precision is obtained in $1-|\lambda|$~\cite{flammia2020efficient},
which is the pertinent quantity to ensure that relative precision is inherited by the Pauli error probabilities we eventually want to estimate.


\subsection{Probability Marginalization}

Pauli eigenvalues are related to Pauli error probabilities through Eq.~\eqref{eq:pauli-eigenvalue}.
Given all Pauli eigenvalues, one only needs to invert Eq.~\eqref{eq:pauli-eigenvalue} to produce the full probability distribution over $n$~qubits. 
However, there are~$|\mathbb{P}_n| =4^n$ distinct Pauli eigenvalues to learn and a corresponding $4^n$ Pauli error probabilities. 
Even for moderate system sizes this becomes impractically many parameters.

That said, many if not most of the error terms in $\mathbb{P}_n$ are unlikely to contribute substantially to the overall error.
A typical and reasonable assumption held in both QEC and QCVV is that high-weight errors are less likely than low-weight errors in many devices. 
With this in mind, rather than considering the full $4^n$ probabilities of the twirled channel, it is more practical to marginalize over the probability distribution and focus on the most important error terms. 

The marginal probability of a Pauli~$P$ over some subset~$S$ of $n$~qubits is defined as
\begin{equation}
    \mu_S(P) = \sum_{\substack{P_i \in \mathbb{P}_n \\  [P_{i}]_S = P}} p(P_i),
    \label{eq:marg-prob}
\end{equation}
where we remind the reader that $P_{i}$ is an $n$-qubit Pauli operator in $\mathbb{P}_n$, and $[P_{i}]_S = P$ denotes that the restriction of $P_i$ to the subset $S$ must equal $P$.
Generically we refer to a marginal probability distribution over $n$ qubits as `$n$-qubit' marginals.

It is reasonable to assume the most substantial error terms will be supported on the qubits on which physical gates are being applied, in our case, the qubits that are being acted upon by \textsc{cnot}s. 
Accordingly, these qubits are the most natural choice over which to marginalize. 
We refer to 2-qubit marginals over a single \cnot as a `1-\textsc{cnot}' marginal distribution. 
Since we are interested in crosstalk we also want to characterize errors that are supported on idling qubits in the register or across multiple \textsc{cnot}s. 
For the latter, we will look at 4-qubit marginal distributions across pairs of \textsc{cnot}s, which we refer to as `2-\textsc{cnot}' marginal distributions.

Marginalizing in this way focuses on relevant error terms, while also drastically cutting down the number of eigenvalues that we need to learn.
As mentioned previously, we also reduce resource costs by using probabilities of error terms associated with a Pauli orbit~\cite{carignandugas2023error} 
\begin{equation}
    \mu_S(P^{\circlearrowright\mathcal{H}}) \coloneqq \sum_{Q\in P^{\circlearrowright\mathcal{H}}}\sum_{\substack{P_i \in \mathbb{P}_n \\ [P_i]_S = Q}} p(P_i)
    \label{eq:full-marg-prob}
\end{equation}
rather than keeping individual Pauli error probabilities. Combining Eqs.~\eqref{eq:pauli-eigenvalue} and~\eqref{eq:full-marg-prob} we can obtain the marginal probabilities of the Pauli orbitals of~$\mathcal{H}$ with the orbital eigenvalues of Eq.~\eqref{eq:orbit-eigval}~\cite{carignandugas2023error}
\begin{equation}
    \mu_S(P^{\circlearrowright \mathcal{H}}) = \frac{|P^{\circlearrowright \mathcal{H}}|}{|\mathbb{P}_{S}|} \sum_{P_i\in \mathbb{P}_{S}} (-1)^{\omega(P,P_i)}\lambda_{P_i^{\circlearrowright \mathcal{H}}}
    \label{eq:marg-WHT}
\end{equation}
for Pauli orbit $P^{\circlearrowright \mathcal{H}} \subset \mathbb{P}_S$. 

By collating marginal probabilities and unique eigenvalues into vectors, $\bm{\mu}$ and $\bm{\lambda}$, we can straightforwardly express equations like Eqs.~\eqref{eq:pauli-eigenvalue} and~\eqref{eq:marg-WHT} in the form $\bm{\mu} = W \bm{\lambda}$. 
If $\bm{\lambda}$ consists of all Pauli eigenvalues in $\mathbb{P}_n$,~$W$ is the Walsh-Hadamard Transform.
More generally, Eq.~\eqref{eq:marg-WHT} allows for~$W$ to have elements that are zero, and may cause certain eigenvalues to contribute asymmetrically to different marginal probabilities.

Estimates of marginalized probability distributions of this kind will often calculate negative probabilities, which are unphysical. 
Negative estimates can sometimes be attributed to insufficient sampling causing poor precision.
However, even adequately sampled estimates can maintain negative probabilities whenever the true probability is close to 0; any amount of error in $\bm{\lambda}$ can immediately incur negativity in the estimated probability. 
It is standard practice to take such probabilities and map them to the nearest all-positive probability distribution on the probability simplex, which generally provides better estimates of the true probabilities~\cite{harper2020efficient,flammia2020efficient}.
That is, find $\bm{\mu}'$ such that
\begin{equation}
    \min_{\bm{\mu}'}\lVert\bm{\mu}' -\bm{\mu}\rVert_2 \text{ s.t. } \bm{\mu}' \geq 0,\quad \bm{1}^T\bm{\mu}'=1.
    \label{eq:prob-proj}
\end{equation}

However, our experimental estimates are for~$\bm{\lambda}$, not~$\bm{\mu}$.
Ideally we should aim to find $\bm{\lambda}'$ close to the estimated $\bm{\lambda}$ such that $\bm{\mu}'$ is positive, that is,

\begin{equation}
    \min_{\bm{\lambda}'} \lVert\bm{\lambda}' -\bm{\lambda}\rVert_2 \text{ s.t. }  \bm{\mu}'=W\bm{\lambda}' \geq 0, \quad 1\geq \bm{\lambda}' >0
    \label{eq:lambda-proj}
\end{equation}
The final condition that $\bm{1}^T\bm{\mu}'=1$ corresponds to $\lambda_{I}=1$, which is always true for trace-preserving maps.

In the case where $W$ is the Walsh-Hadamard Transform, $W^T W \sim I$ such that Eqs.~\eqref{eq:prob-proj} and~\eqref{eq:lambda-proj} are entirely equivalent. In our case, since we marginalize over Pauli errors in the same orbit, we have $W^T W \not\sim I$. Accordingly we will use Eq.~\eqref{eq:lambda-proj}, which takes into account how some eigenvalues have a larger impact on particular marginal probabilities.

\subsection{Graphical model under Gibbs random field}
\label{sec:GRF}

As the number of physical qubits becomes large, even simply describing the full probability distribution of Pauli error rates becomes inefficient.
We use the concept of the \textit{Gibbs Random Field} (GRF)~\cite{flammia2020efficient,harper2020efficient}, which is a strictly positive probability distribution obeying certain physically motivated assumptions known as \textit{global Markov property}. The global Markov property is defined as follows: assume that for two variables $x_A$ and $x_B$, over qubits $A$ and $B$, there is a separating set of vertices $S_{A,B}$ (and these vertices are associated with variables $x_{S_{A,B}}$), that is, for every path from $A$ to $B$ we go through at least one vertex in $S$. Then the global Markov property asserts that $p(x_A,x_B|x_{S_{A,B}})=p(x_A|x_{S_{A,B}})p(x_B|x_{S_{A,B}})$, for every choice of variables $x_A$ and $x_B$ and separating set $S_{A,B}$. In other words the marginal distribution over $A$ and $B$ is conditionally independent given the values of the separating set variables. This can be graphically displayed in a factor graph, which has two types of nodes: variable nodes which are qubits and factor nodes which connect these variables. These factor nodes describe the correlations that are possible according to the global Markov property. Importantly the Hammersley-Clifford theorem \cite{Hammersley1971} states that if a strictly positive distribution obeys the global Markov property for a factor graph, it can be factorized over that graph.

Proximity in a factor graph does not necessarily imply physical proximity.
It rather implies a mechanism to spread correlation that necessitates some form of coupling, which may depend on proximity in some parameter including space or frequency.
There are many different mechanisms by which error correlations can spread.
A prime avenue for the introduction of correlated errors is via the control to which we subject the computational system.
Trapped-ions are comparatively well isolated and therefore suffer relatively few common sources of correlated errors.
One such source is optical crosstalk whose strength typically decays with physical distance such that error correlations are predominantly nearest-neighbor while next-nearest-neighbor is usually already below threshold for detection.

The entangling gate mechanism itself utilizes the motional modes of the ion chain which are global properties of the crystal and are therefore prominent candidates to spread correlations non-locally.
The avenue by which error correlations could spread is residual spin-motional entanglement at the end of a gate.
Such residual entanglement can arise from miscalibration or stochastic fluctuations in gate parameters.
The sensitivity of the gate to these effects depends further on the average phonon occupation in the mode(s) of interest.
Under ideal conditions, the gate will address only two ions out of the full ion register, leaving the remaining qubits in a product state between spin and motion.
Upon completion of an imperfect gate, residual correlations between the motional and electronic state remain for the ions that were interacted with, while the ions in the rest of the register are unaffected.
As we discussed before, optical crosstalk that would introduce entanglement between spin and motion on the remaining register is limited predominantly to nearest-neighbor interactions.

The error correlations of single-qubit and entangling gates realized in the used trapped-ion experiment are assumed to be nearest-neighbor. We can abstract to a Gibbs random field connectivity graph as discussed in Sec.~\ref{sec:Results}, which we will use to make predictions about logical performance. Each qubit is connected to its partner in the entangling gate and the corresponding neighbors.

The considerations above are tailored to our experiment and may not hold in general. Other platforms may be dominated by different error sources and error correlations, and the details of the relevant GRF model and its compactness will vary. Regardless, this GRF framework can be applied generally.


\subsection{Optimizing Experimental Parameters}
\label{sec:Optimization}

The time budget in many QCVV settings is primarily determined by the number of discrete configurations that must be run, which includes the desired number of distinct sequence lengths, initial states, and measurement bases.
In our experimental context, the number of Pauli-frame randomizations multiplies the number of distinct configurations as well for the purpose of budgeting time.
The number of configurations ultimately increases with the number of unique eigenvalues that have to be learned in CER.
Therefore, concentrating on marginal probabilities greatly reduces the number of experimental parameters required.
The exponential scaling of the number of eigenvalues with qubit number~$n$ reduces to an order-$k$ polynomial when focusing on all possible marginalizations over $k$~qubits since \mbox{$4^k\binom{n}{k} = O(n^k)$}.
Here we marginalize over single \cnots and pairs of \textsc{cnot}s, corresponding to~$k=2$ and~$k=4$, respectively.

In the same vein, we want to learn chosen experimental parameters in as few experimental runs as possible. 
One can extract up to $2^n$~different Pauli expectation values in parallel, but generally this requires entangled input states and measurements which contribute to SPAM. 
However, for hard cycles from the Clifford group certain finite sequence lengths are equivalent to performing the identity operation.
Our hard cycle is a transversal application of \textsc{cnot}s, so we specifically have $\mathcal{H}^2(\rho) = \rho$. 
Hence, by choosing $\mathcal{H}^m$ such that $m$ is even, we guarantee that easily-prepared Pauli product states will be mapped to Pauli product measurements, keeping the contribution of SPAM low.
At the same time, the measurements resolve each qubit and we can infer multi-qubit Pauli measurement outcomes by taking the product of single-qubit Pauli measurement outcomes. This reduces the number of experiments further because we are able to extract all $2^n$~distinct Pauli expectation values from $n$-qubit product states.

\begin{figure}
\centering
\includegraphics[width = 8.5cm]{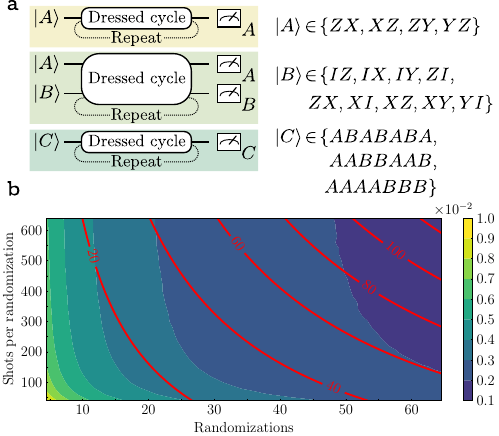}
\caption{Optimization of experimental resources.
\textbf{a}~The number of distinct experiments to run for a single, two, and seven parallel \cnots to obtain all unique orbits is 4 (the eigenstates of elements from the set $A$, for 1-\cnot orbits), $4\times9 = 36$ (all combinations of $A$ and $B$, for 2-\cnot orbits), and $36\times3 = 108$ (all tilings $C$ of elements from $A$ and $B$, for all possible 2-\cnot orbits over 7 \textsc{cnot}s), respectively.
\textbf{b}~Tradeoff between number of randomizations per initial state and shots per randomization for an eigenvalue estimate of $\lambda_{ZX}$.
The color scale depicts the standard deviation in estimates of $\lambda_{ZX}$ at the chosen experimental parameters, estimated by subsampling over an experimental data set.
In total the data set included 300 randomizations with 650 shots each.
In our platform changing the randomization is more costly than repeating a shot.
Time in minutes to estimate a fixed initial state using three values for~$m$ is indicated by red contours.
Total shots required is obtainable by the product of the shots per randomization, total randomizations, the distinct sequence lengths $m$, and the distinct initial states.}
\label{fig:Optimization}
\end{figure}

Expounding on our particular case, we can determine the specific number of configurations that we require, as illustrated in Fig.~\ref{fig:Optimization} \textbf{a}. 
By choosing even values for~$m$ and only preparing states that give unique orbital eigenvalues we can measure all 1-\textsc{cnot} orbital eigenvalues (63 non-trivial eigenvalues in total) with 4 initial states.
Notably, the number of initial states required for 1-\cnot marginals remains the same for an arbitrary number of \cnots applied in parallel.
The reason this is achievable is because preparing Pauli eigenstates on one \cnot does not obstruct the preparation of Pauli eigenstates on another \textsc{cnot}, that is, the same initial state can be prepared on each respective \cnot in parallel. 
However, the same is not possible for 2-\textsc{cnot} marginalization, where some 4-qubit marginals may preclude another 4-qubit marginal from being measured at the same time. 

A naive estimate for the total number of configurations needed for 2-\cnot marginals would be $4\times4=16$ initial states. However, the set $(P_1P_2)^{\circlearrowright \mathcal{H}} \neq P_1^{\circlearrowright\mathcal{H}} \times P_2^{\circlearrowright\mathcal{H}}$. This can be observed simply by noting the cardinality of the sets: $(P_1P_2)^{\circlearrowright \mathcal{H}} = \{P_1P_2,\mathcal{H}(P_1P_2) \}$ has an orbital length of two but $P_1^{\circlearrowright\mathcal{H}} \times P_2^{\circlearrowright\mathcal{H}}$ is a set of 4 elements. For our choice of hard cycle, equality is satisfied whenever $P_1$ or~$P_2$ is mapped to itself, up to a change in sign. 
Consequently we often need to prepare $P_1$ \textit{and} $P_1P_2$, increasing the number of initial states from 16 to 36. 
When considering all possible 2-\cnot marginalizations in the transversal seven \cnot case, three distinct tilings, shown in Fig.~\ref{fig:Optimization}~\textbf{a}, are needed such that 100 initial states are required in total ($3\times 36 = 108$, minus 8 redundancies) to learn \mbox{$4^4\times \binom{7}{2} = 5376$} expectation values.

Aside from the number of distinct configurations that must be measured, another key factor in the overall resource budget is how many shots have to be allocated to each configuration. 
The precision with which the decay parameters can be estimated depends on the number of samples, due to shot noise. 
At the same time, randomized compiling requires averaging over a sufficiently large number of Pauli-frame randomized circuits in order to achieve meaningful decays. 
Ideally one would measure each shot allocated to a particular eigenvalue estimation with a different instance of random interleaved Pauli operators. 
However for many hardware platforms, including ours, different configurations -- irrespective of whether they are required by randomized compiling or to obtain different eigenvalues -- are more costly than repeated execution of the same experiment.

There is therefore a tradeoff between experiment wall-clock time and estimator variances, governed by how many shots are performed per randomization and the number of randomizations in total for a fixed overall budget.
Some platforms admit hardware-assisted execution of parametrized circuits~\cite{rajagopala2024hardware} and hardware-efficient randomized compiling~\cite{fruitwala2024hardware} that reduce or entirely eliminates this additional time overhead.
Similarly, we can note that through Eq.~\eqref{eq:marg-WHT} some eigenvalues are more important in the determination of the marginal error probabilities than others; preferential shot allocation to these can improve estimation precision at fixed budget as well.

For each experiment the results of Fig.~\ref{fig:Optimization}~\textbf{b} were used as a rough guideline for the amount of time that would be necessary to achieve some desired level of accuracy, and what parameters would be appropriate to doing so.
We typically used 40 randomizations per initial state and 150 shots per randomization for measuring 1-\cnot orbital eigenvalues and minimize the variance. One initial state took around 40 minutes.
The 2-\cnot orbital eigenvalues were typically measured with 20 randomizations and 200 shots per randomization, increasing the variance slightly and reducing the run time per initial state to roughly 20 minutes.

\subsection{Uncertainties and noise floors}
\label{sec:uncertainties}

\begin{figure*}[t]
\centering
\includegraphics[width = 17cm]{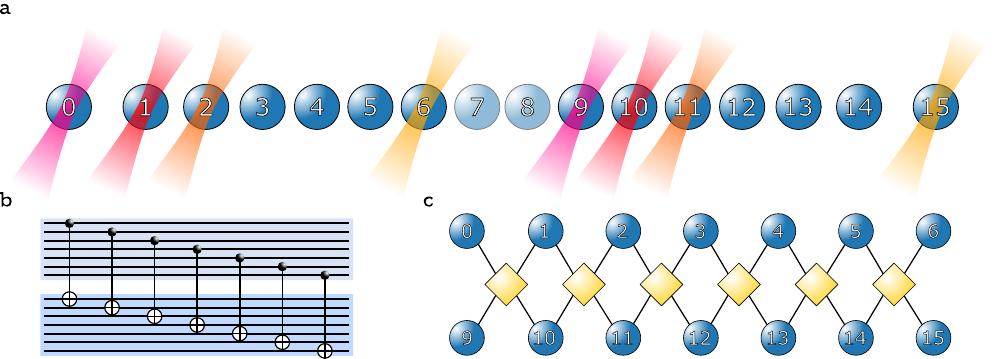}
\caption{Arrangement of a transversal \cnot on a 16-ion register at different levels of abstraction. \textbf{a} An ion chain in a harmonic potential arranges with inter-ion distances growing towards the edges. Pairs of tightly focused laser beams (color gradients of same shade) are used to address ion pairs. Ions at the center are expected to see more crosstalk from neighbors due to tighter packing. Ions 7 and 8 (lighter shade) always idle. \textbf{b} Circuit layer abstraction of the chain with two qubit registers that are logically separate (shading) and are connected via a transversal \cnot only. \textbf{c} Gibbs random field abstraction where qubits (blue spheres) possess error correlations only with others if they are connected via nodes in the graph (yellow squares).}
\label{fig:Transversal_CNOT_Setup}
\end{figure*}

There are many factors that are culpable for how estimated quantities can vary between experiments.
Some statistical factors are anticipated as part of the experimental design and therefore entirely controllable, such as shot noise and randomizations (see Sec.~\ref{sec:Optimization}).
However, there are also factors that are outside of the control of an experiment, for instance in the form of device drift.
That being the case, it is important to contextualize all estimates with error bars, ranges and bounds.

Error bars shown in figures depicting marginal error rates are determined by bootstrapping the data over the time period of the experiment.
Where calculated, the range is reported to \textit{one} standard error unless stated otherwise.
Such bars do not bound the long-term performance of the device, rather, they signify the variance in the experimental data over the time-frame that the data was obtained.

Under the assumption that errors on the easy cycles are gate-independent, which includes the case where easy gates have negligible error, the CER theory is exact.
In this case errors estimated under CER correspond to the composition of the (arbitrary) error map on the hard gate cycle of interest combined with the error map from the easy gate cycles \cite{wallman2016noise,carignandugas2023error}.
If the easy gate errors exhibit significant gate-dependent variation, then the error reconstruction from CER is no longer exact, but has a correction term (inherited from the theory of randomized compiling~\cite{wallman2016noise}) that decreases quadratically with the average variation over the errors across the easy gate set~\cite{carignandugas2023error}.
This variation is upper-bounded by the average error rate over the set of easy gates \cite{wallman2016noise}, which can be independently determined by standard randomized benchmarking \cite{magesan2011scalable} or cycle benchmarking \cite{erhard2019characterizing}.

In our experiment, the easy cycles are not entirely noiseless and thus confer an average-noise channel of their own for each hard cycle. 
Since each single-qubit operation is applied serially, the remaining qubits idle as one qubit is acted upon by a random Pauli operation.
Over an entire easy cycle dephasing errors can build up in addition to noise from imperfect $X$ and $Y$ pulses.
The easy cycle error rate is an order of magnitude smaller than the hard cycle error rate, however, the total error rate of the average-noise channel of the easy cycle is the same order of magnitude as the marginal error rates of \textit{specific} Pauli errors.

Experiments to ascertain the overall error rates of the average-noise channel were conducted by applying a series of random Pauli cycles of different sequence lengths and marginalizing over individual qubits in the register.
The marginal error rate of the average-noise channel was determined for each qubit and these were averaged to give the average single-qubit marginal error rate of the average-noise channel.
The value of this single-qubit marginal error rate has been included as a noise floor in figures presenting marginal error rates, illustrated in light-gray.
A dark-gray region is also shown at one third of the light-gray value, which would be the error rate of a specific Pauli error assuming depolarizing noise from the easy cycle.
Error terms that are within the light-gray region are likely to be significant errors, but deserve scrutiny.


\section{Experimental setup}
\label{sec:Experimental}

All experimental results presented in this manuscript are implemented in a trapped-ion quantum processor~\cite{pogorelov2021, postler2022, postler2024demonstration}. Sixteen $^{40}$\textrm{Ca}$^+$ ions are trapped in a macroscopic linear (`blade') Paul trap with asymmetric RF drive configuration. The trap parameters are set such the ions crystallize in a linear chain. The motional and electronic states of the ions are controlled via laser and radio-frequency pulses. Each ion encodes one qubit in the electronic Zeeman sublevels \mbox{$\ket{0}=\ket{4^2\textrm{S}_{1/2}, m_J = -1/2}$} and \mbox{$\ket{1}=\ket{3^2\textrm{D}_{5/2}, m_J = -1/2}$} connected via an optical quadrupole transition at a wavelength of \SI{729}{\nano\meter}.

The Coulomb interaction between the ions gives rise to collective motional modes of the ions, which are used to mediate entangling operations between any desired pair of qubits. Single-ion addressing in the radial chain direction with light at the \SI{729}{\nano\meter} qubit transition allows for arbitrary single-qubit and arbitrary-pair two-qubit gate operations. The harmonic potential confining the ions leads to inter-ion spacings that are non-uniform such that addressing errors like optical crosstalk depend on which ions are addressed, as seen in Fig.~\ref{fig:Transversal_CNOT_Setup} \textbf{a}.
More details on the setup can be found in Ref.~\cite{pogorelov2021}.

The native gate set used in the setup consists of arbitrary-angle single-qubit rotations $R(\theta, \phi)$ around any axis in the equatorial plane of the Bloch sphere (pulse area $\theta$ and angle $\phi$ relative to $+X$) implemented via laser pulses resonant with the optical qubit transition, single-qubits $Z$-rotations `virtually' implemented in software, and maximally-entangling two-qubit gates \mbox{$MS(\pm\pi/2) = \mathrm{exp}(\mp i\frac{\pi}{4} X\otimes X)$} implemented via the \text{M\o{}lmer-S\o{}rensen} interaction~\cite{sorensen1999}. The $MS$ gate belongs to the class of geometric phase gates generated by optical forces where entanglement between qubits not natively interacting is created via a mediating boson; here excitations of the joint motional mode of the ion crystal. The $MS(\pm\pi/2)$ gate is equivalent to the \cnot gate up to local rotations~\cite{maslov2017}, compiled here as
\begin{equation}
\label{eq:CNOT_compilation}
\includegraphics[width=8.5cm]{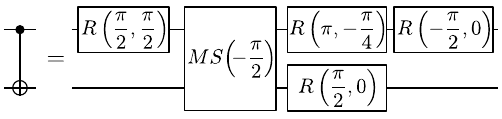}
\end{equation} 
The $MS$ gates employed here utilize the center-of-mass radial motional mode of the ion crystal as the mediating boson. Using the global phononic bus of the motion to create entanglement means that we have to perform $MS$ gates serially even when two gates feature disjoint support.
There are control-theoretic gate modulation techniques to circumvent this limitation, e.g. in Ref.~\cite{figgatt2019parallel}, which are however not part of this work.

All experiments are submitted as Qiskit~\cite{qiskit2024} circuits that undergo a transpilation procedure contracting single-qubit gates before and after a hard cycle shown in Fig.~\ref{fig:pauli-eigenvalue-learning}. The setup is subject to continuous, automatic calibration routines, whose purpose is to maintain system performance. These routines autonomously halt user-queued measurements and adjust experimental parameters before resuming.
A consequence of this is that any measurement will be subject to drifts and stochastic fluctuations, as well as discontinuous changes in operational parameters. Characterization experiments therefore have to be understood as long-term average machine behavior, which we argue is both extremely common and the most reasonable interpretation.

For the purposes of this study we map qubits to ions as shown in Fig.~\ref{fig:Transversal_CNOT_Setup} \textbf{a}. Two logical qubits, each encoded in a 7-qubit code, requiring $2\times7$ code qubits; ions 7 and 8 are chosen to idle since errors are expected to be highest at the center of the chain.
The choice of mapping from ions to qubits is non-unique and different mappings present tradeoffs for certain implementation aspects. The ion-to-qubit mapping chosen here maximizes average distance between the ions involved in individual \cnots of the transversal setup shown in Fig.~\ref{fig:Transversal_CNOT_Setup} \textbf{b} but is not assumed to be optimal in any respect.


\section{Results}
\label{sec:Results}

In this section, we present the analysis of CER applied to individual \textsc{cnot}s, as well as to seven \cnots acting transversally as a logical operation.

We analyze how the noise profile of an individual \cnot changes in different experimental contexts, that is, as we vary which qubits in the ion chain the \cnot acts and the size of the hard cycle.
When operated transversally, we find with a fixed qubit-to-ion mapping that our transversal \cnot performance is well described by the individual \cnots (component-level) in conjunction with simple device physics: additional dephasing through idling.

We further estimate logical error rates from physical errors characterized up to 2-\cnot marginal distributions using the nearest-neighbor Gibbs random field model of Fig.\ref{fig:Transversal_CNOT_Setup}~\textbf{c}.
This particular choice of simple model is based on the observation that error correlations in the ion-trap platform are typically only short range, which is corroborated by the 2-\cnot characterization afforded by CER. 


\subsection{Error characterization of a single CNOT}
\label{sec:Single_CNOT}

We begin by studying how the noise profile of a \cnot depends on its support in the ion crystal.
The laser beams used to control the ions' electronic and motional state have finite spatial extent even in the absence of optical imperfections, meaning that optical crosstalk depends on distance between ions.
The spacing of ions in the crystal as formed in a harmonic potential differs at the center and the periphery of the chain.
Ions will also experience increasing levels of micromotion the further they are away from the trap potential minimum.
The \text{M\o{}lmer-S\o{}rensen} gate implemented in this work uses primarily the radial center-of-mass motional mode of the ion chain to mediate entanglement, whose coupling strength is uniform across the chain.
However, close-by higher motional modes contribute to the gate as well, whose coupling and contribution to errors depends on an ion pair's position in the chain.
Combining these, we expect errors that arise through the drive to depend on the support of the gate, even assuming otherwise ideal behavior. There are many different experimental imperfections that will also depend on ion position.

\begin{figure}[t]
\centering
\includegraphics[width=8.5cm]{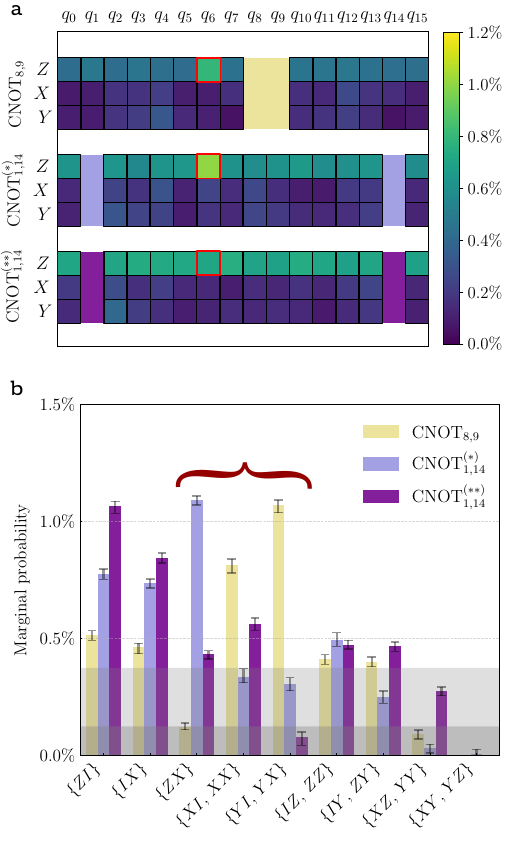}
\caption{Noise profile of a single \cnot embedded in a 16-ion register over different supports. 
\textbf{a} Heatmap of single-qubit marginal error rates over each idling ion for three distinct single \cnot experiments. Qubits that the \cnot is supported on are colored as in \textbf{b}. Red boxes highlight, before and after removal, a high $Z$ error on qubit 6 caused by incorrect quadrupole shift compensation.
\textbf{b} Comparison of 2-qubit marginal error rates on the single \textsc{cnot}.
Braces highlight error terms that are notably distinct across different supports, in particular the dominant error rates for \textsc{cnot}$_{8,9}$ are distinct from those found for \textsc{cnot}$_{1,14}$ which acts on non-adjacent qubits.
We find that $ZX$ error terms are often associated with miscalibrations of gate power, as highlighted in Fig.~\ref{fig:Seven CNOT}. 
A dominant $ZX$ error term on \textsc{cnot}$^{(*)}_{1,14}$ is reduced in the later experiment \textsc{cnot}$^{(**)}_{1,14}$.}
\label{fig:miscal}
\end{figure}

We now demonstrate the use of CER for characterizing individual physical \cnot on two different supports, qubit pairs (1, 14) and (7, 8).
As shown in Fig.~\ref{fig:miscal} \textbf{a}, idling qubits see predominantly $Z$ errors, which arise through uniform dephasing caused by magnetic field fluctuations on spatial extents much larger than the ion chain, as well as laser frequency and phase fluctuations that are also homogeneous across the register.
As shown in Fig.~\ref{fig:miscal} \textbf{b}, for the 1-\textsc{cnot} error terms we can generally see a hierarchy of orbits for the magnitude of the associated error: Orbits with Pauli weight-1 such as $ZI$ have the highest errors, followed by orbits that have mixed weight-1 and weight-2 such as $\left\{XI,XX\right\}$, and orbits of only weight-2 such as $\left\{YZ,XY\right\}$ being the lowest.
This is the ordering we expect given the expectation that individual Pauli errors are weakly correlated.
Deviations from this ordering, more specifically higher order terms deviating from simply the product of individual rates, is indicative of quantum-correlated errors.
When the support of the \cnot is on ions that are directly adjacent to each other such deviation can be observed, where the mixed-weight orbits $\left\{XI,XX\right\}$ and $\left\{YI,YX\right\}$ are much larger on \textsc{cnot}$_{8,9}$ than their counterparts on \textsc{cnot}$_{1,14}$.
On the other hand, the qualitative error profile for qubits that are further separated are similar.

Discerning different errors on individual qubits provides opportunities to identify and remove errors at their source, as we now demonstrate.
The \cnots in our setup are implemented using the compilation into the native gate set in Eq.~\eqref{eq:CNOT_compilation}.
This means that on a circuit level the information gathered about \cnots is useful but it obfuscates the origin.
For example, the increase in $XX$ errors for adjacent support, shown in Fig.~\ref{fig:miscal} \textbf{b}, could be explained by a $ZI$ error just before the $MS$ gate.
This error is more likely for adjacent support since the addressed interaction laser beams have high intensity and cause AC Stark shifts at neighboring ions; a coherent effect that is only evident in this configuration.

Other error sources may enter through miscalibration and can be removed upon identification, as we show in Fig.~\ref{fig:miscal} \textbf{a}.
While the majority of idling qubits suffer near-identical dephasing errors, qubit 6 is more affected.
There are few spatially-dependent effects that would cause dephasing so strongly localized.
Using the fine-grained information obtainable in CER we identified qubit 6 to have a faulty setting for the optical quadrupole shift compensation we apply to all laser pulses.
Similarly, the weight-2 orbit $\left\{ZX\right\}$ has by far the largest marginal error rate, even though we would expect the opposite based on independent error sources.
This is likewise observed in Fig.~\ref{fig:Seven CNOT} \textbf{a}.
Through the \cnot compilation we however know this error to be a correlated $XX$ error at the native gate level, which may be caused by a gate power miscalibration (see App.~\ref{sec:ZX_Errors} for details).
Knowing this, we tracked the first occurrence of this behavior in time and could ultimately trace it to an incorrect setting of the applied laser power depending on the ion pair.
We stress that both of these error sources have been inadvertently introduced and were subsequently easily removed, but their identification using CER reinforces the diagnostic utility that it provides, directing the search for error sources rather than simply stating the existence of more or fewer errors.

The overall wall-clock runtime for CER of a single \cnot suite with 40 randomizations per initial state and 150 shots per randomization on one support takes about 2-3 hours after optimization; this time includes automatic calibration routines and other overhead of operation.


\subsection{Transversal \textsc{CNOT}}

\begin{figure*}[t]
\includegraphics[width = 17cm]{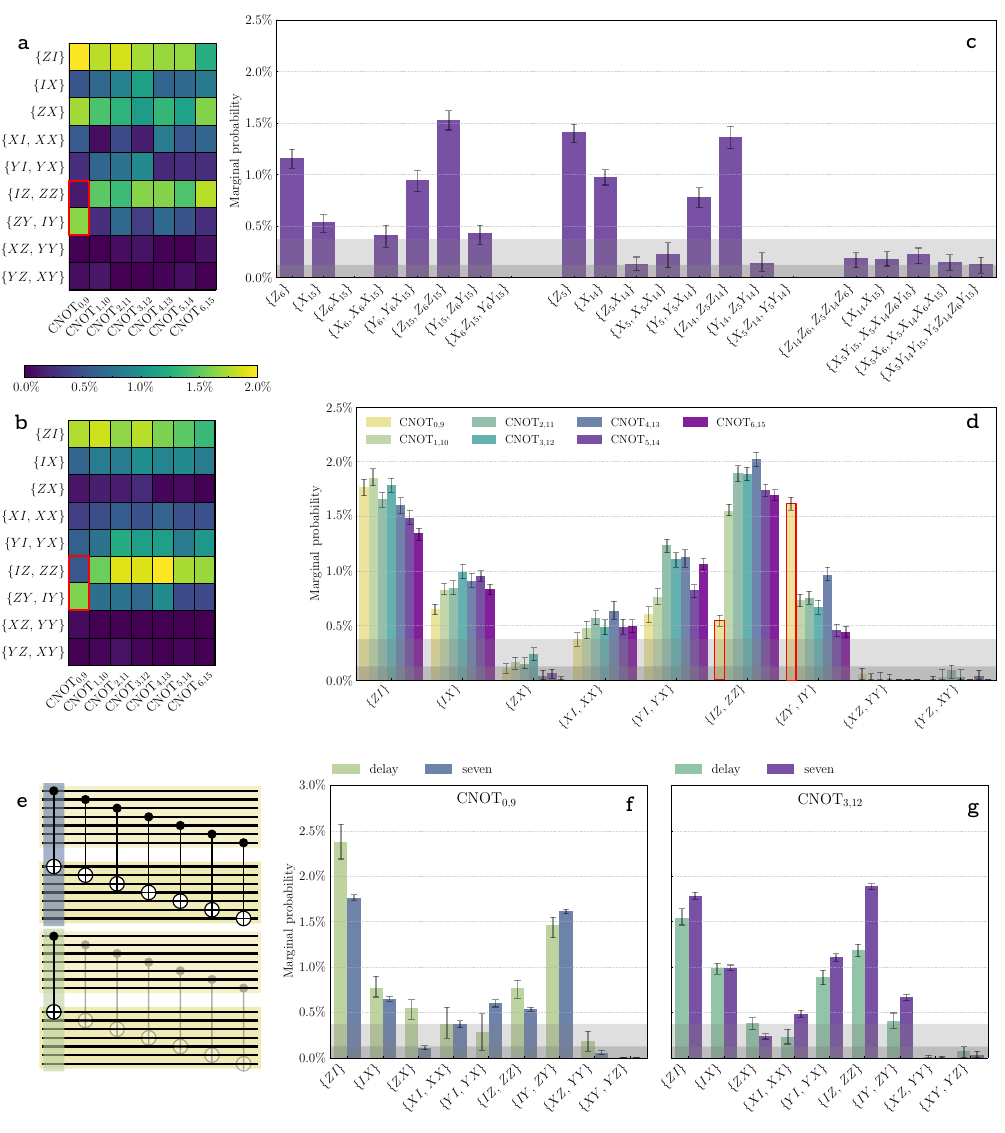}
\caption{Characterizations of a 14-qubit transversal \cnot $\mathcal{H} = \prod_{i=0}^6\textsc{cnot}_{i,i+9}$ in a 16-ion register.
\textbf{a}~Heatmap display of 1-\textsc{cnot} marginals for each constituent \textsc{cnot}. The $ZX$ error terms are identified to be caused by a miscalibration of the $MS$ gate power.
\textbf{b}~A heatmap display of 1-\textsc{cnot} marginals for a separate characterization experiment, after the source of the $ZX$ signal was identified and eliminated. 
In \textbf{a} and \textbf{b}, marginals $\mu_S(ZI^{\circlearrowright\mathcal{H}})$ and $\mu_S(IZ^{\circlearrowright\mathcal{H}})$ associated with dephasing are larger than for a single \cnot in isolation. The error characteristics for each of the individual \cnots is similar (little change horizontally) with the exception of \textsc{cnot}$_{0,9}$ for $\mu_S(IY^{\circlearrowright\mathcal{H}})$ and $\mu_S(IZ^{\circlearrowright\mathcal{H}})$ (highlighted in red).
\textbf{c}~Significant error terms for a 2-\cnot marginal, taken at the same time as \textbf{b}, over \textsc{cnot}$_{5,14}$ and \textsc{cnot}$_{6,15}$. Marginal terms only supported on $(6,15)$ are shown first, then terms only supported on $(5,14)$, and then all other terms with error rates higher than the dark-gray region.
\textbf{d}~The same seven orbits from \textbf{b} as histograms, with the anomalous strengths of \textsc{cnot}$_{0,9}$ again highlighted in red. Apart from the anomalous terms in \textsc{cnot}$_{0,9}$, there is a much higher amount of dephasing on control and target qubits due to time spent idling. Since the predominant difference from single \cnots in isolation appears to be idling, we compare the noise profile of individual \cnots from the transversal \cnot of \textbf{d} with a single \cnot undergoing an equivalent period of idling.
\textbf{e} Experimental sequence for labels `delay' and `seven' in \textbf{f}. The top circuit executes seven \cnots successively and we study the marginal on support $(0,9)$ and $(3,12)$. The bottom circuit executes a single \cnot but idles for an additional time equivalent to 6 \cnots being applied in sequence. 
\textbf{f} and \textbf{g} are histograms of the marginals over \textsc{cnot}$_{0,9}$ and \textsc{cnot}$_{3,12}$, respectively, either as part of the full transversal \cnot or as a single \cnot with an equivalent idling time of the transversal \textsc{cnot}. 
Data in \textbf{g} was taken with machine performance deteriorated before ongoing larger maintenance.
The idling errors of \textbf{f} resemble the transversal \cnot errors, rather than additional dephasing.
}
\label{fig:Seven CNOT}
\end{figure*}

We now apply CER to the transversal \cnot as would be applied in many Calderbank-Shor-Steane (CSS) codes, including the 7-qubit Steane code.
A transversal \cnot here means that a logical operation between two code blocks is executed by applying bit-wise physical \cnots between the constituent physical qubits of each code.  Transversal operations are ideal for performing quantum logic, as they restrict the propagation of errors.
The transversal \cnot also satisfies our criterion for a hard cycle since the supports of individual operations are disjoint.
In the all-to-all connectivity of trapped-ions we are free to choose this mapping in whatever way we deem appropriate; a separation of code blocks, for example, does not have to be at all reflected in physical locations.
The mapping we chose to adopt here, however, is very similar to the abstract arrangement of qubits in the code, as shown in Fig.~\ref{fig:Transversal_CNOT_Setup} \textbf{b}.
This is a choice that follows from our previous considerations about spatially-dependent error sources -- in particular optical crosstalk -- and is backed by our findings for single \textsc{cnot}s.

We expect the results of CER applied to the transversal \cnot will differ from the individual single \cnot results performed separately.
When we execute several \cnots in a hard cycle, each individual \cnot now has neighboring qubits that have been or will be acted upon by non-trivial entangling operations. 
Additionally, any particular component \cnot in the set of seven \cnots has to idle for an amount of time equivalent to six more \cnots before it will be applied again in the next step of the hard cycle repetition.
This is due to the fact that our entangling gates have to be applied serially, rather than in parallel, through the gate mechanism.
We would expect therefore more crosstalk avenues and larger error contributions caused by dephasing, leading to overall higher error rates.
The choice of ion-to-qubit mapping here means that we would expect similar noise profile on all pairs, with some variation due to the asymmetry of the \cnot itself, as well as potential deviations for \cnots at the edge of the chain because they only have one neighbor rather than two.

We perform CER on a transversal \cnot realized as in Fig.~\ref{fig:Transversal_CNOT_Setup}.
The noise profile of the individual \cnots is very similar (near-uniform rows in Fig.~\ref{fig:Seven CNOT} \textbf{a} and \textbf{b} or similar-height columns in Fig.~\ref{fig:Seven CNOT} \textbf{d}) with the exception of \textsc{cnot}$_{0,9}$, where the roles of $\mu_S(IY^{\circlearrowright\mathcal{H}})$ and $\mu_S(IZ^{\circlearrowright\mathcal{H}})$ do not match expectations.
We further see that the ordering of marginal sizes is the same as for individual \cnots performed in isolation, that is that the largest marginals contain weight-1 Pauli operators, followed by mixed orbits, and orbits consisting of weight-2 terms only are the smallest.
A departure from this trend is presented by the mixed orbit $\left\{IZ,ZZ\right\}$ (or $\left\{IY,ZY\right\}$ for \textsc{cnot}$_{0,9}$), that is nearly as large as the $ZI$ term.
All of these marginals are associated with single-qubit dephasing errors, likely caused by magnetic field fluctuations in the system.

When we marginalize over a particular \cnot in the transversal setup (single column in Fig.~\ref{fig:Seven CNOT} \textbf{a} and \textbf{b} or single color in Fig.~\ref{fig:Seven CNOT} \textbf{d}) we may compare this with CER performed on the same \cnot in isolation, as explored in Fig.~\ref{fig:Seven CNOT}.
We see that the contributions associated with dephasing in the transversal \cnot ($\{ZI\}$ and $\{IZ, ZZ\}$) are the largest and are far greater than would be expected based on the single \cnot experiment.
However, this can be realigned with expectations if we recall that the overall idling time per \cnot in the transversal setup is substantially larger than for a single \cnot where idling is only over the course of a single easy cycle, rather than an idling time equivalent to six \cnots in sequence.
Consequently, we compare the 1-out-of-7 \cnot data to a single \cnot where we add an equivalent waiting time, the sequence being schematically shown in Fig.~\ref{fig:Seven CNOT} \textbf{e}, data in Fig.~\ref{fig:Seven CNOT} \textbf{f} and~\ref{fig:Seven CNOT} \textbf{g} in green and blue, respectively.
With both \cnots idling a similar amount we can see good agreement, noting also that these different data sets were taken considerable time apart from each other leading to slightly different experiment performance.

The overall wall-clock runtime for CER of a transversal \cnot suite with 20 randomizations per initial state and 200 shots per randomization on seven supports takes about 40 hours after optimization; this time includes automatic calibration routines and other overhead of operation.


\subsection{Predicting logical error rates and performance}
\label{sec:PredictionLogicalErrors}

We now consider how the results of our CER experiment and analysis can inform predictions about the performance of the transversal \cnot operated as a logical gate on a pair of quantum codes, as well as predictions about the potential logical performance of such a transversal \cnot if quantum error correction were to be applied.

First, we note that our CER experiment does not provide the entire probability distribution of all Pauli errors, as obtaining this joint probability distribution would require a number of experiments performed that scales exponentially in the system size -- something we aim to avoid by using CER.
Indeed, the resources required to estimate the full distribution are tantamount to those required for full process tomography.

Nonetheless, as we show, our CER data can yield useful and actionable information about physical and logical error characteristics beyond just the few-physical-qubit marginals analyzed above, using a graphical model in the form of a Gibbs random field, as described in Sec.~\ref{sec:GRF}.
We adopt a graphical model described by the factor graph shown in Fig.~\ref{fig:Transversal_CNOT_Setup} \textbf{c} for the characterization of a transversal \textsc{cnot}.
Under this physically motivated nearest neighbor correlation model we only require marginals up to 2-\cnot to determine all other Pauli error probabilities, taking each error in the orbit of a marginal to have equal probability. Using this model, the joint probability distribution of any random variable $x$ is given by
\begin{equation}
    p(x) = \Bigg(\prod_{i=0}^{5} p\left(x_{\{i,i+9\}} \mid x_{S_{\{i,i+9\}}}\right)\Bigg) \times p\left(x_{\{6,15\}}\right)
\end{equation}
and using the definition of conditional probability, we can write it as
\begin{equation}
    p(x) = \Bigg(\prod_{i=0}^{5} \frac{\mu\left(x_{\{i,i+9\}}, x_{S_{\{i,i+9\}}}\right)}{\mu\left(x_{S_{\{i,i+9\}}}\right)}\Bigg) \times \mu\left(x_{\{6,15\}}\right)
    \label{eq:joint prob}
\end{equation}
where $S_{\{i,i+9\}}$ is the separating set for qubits $i$ and $i+9$, which is the nearest neighbors of $\{i,i+9\}$, serving as adjacent support.
\begin{figure}
    \centering
    \includegraphics[width=8.5cm]{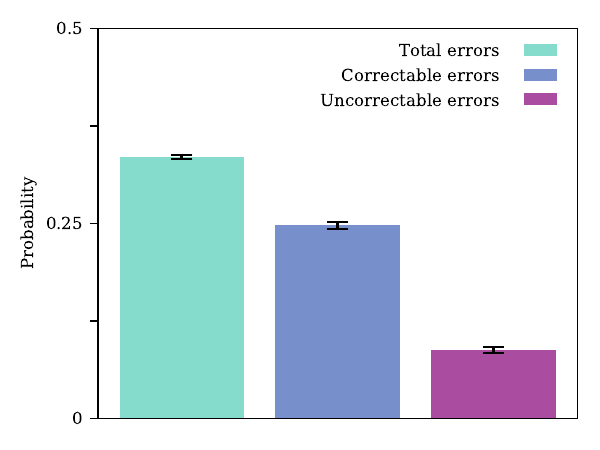}
    \caption{Probabilities of different error scenarios reconstructed from transversal \cnot CER data using the GRF factor graph of Fig.~\ref{fig:Transversal_CNOT_Setup} \textbf{c}. Reconstructed non-zero error rates are grouped into uncorrectable and correctable errors (including errors stabilizer-equivalent to correctable, see Sec.~\ref{sec:GRF}) assuming two logical qubits encoded in 7-qubit Steane codes connected via a transversal \textsc{cnot}. The uncorrectable error rate has probability $0.088\pm 0.004$, compared to total error rate $0.335\pm 0.003$. Error bars are obtained by bootstrapping the experimental data to generate marginal error rates, as outlined in Sec.~\ref{sec:uncertainties}. The reconstructed rates are reported here to two standard errors.}
    \label{Fig:Logical_Errors}
\end{figure}

Given access to marginals up to 2-\textsc{cnot}, such as shown in Fig.~\ref{fig:Seven CNOT} \textbf{c}, we can reconstruct the joint probability distribution of all random variables across the nodes of the factor graph under the local Markov conditions using Eq.~\eqref{eq:joint prob}.

With this joint probability distribution, we can compute logical error rates associated with the transversal \cnot by classifying errors into correctable and uncorrectable, in accordance with the properties of the Steane code.
The Steane code, having distance $d=3$, can correct one arbitrary weight-1 error on a single logical code block. 
Furthermore, since it is a CSS code it may correct weight-2 errors comprised of $X$ and $Z$, namely $X_i Z_j$ for $i\neq j$.
Here, we look in particular at two copies of the 7-qubit Steane code that are connected via a transversal \textsc{cnot}.
In that context, errors remain correctable so long as they satisfy these conditions independently on either code block.
This means all errors supported on a single \cnot are correctable, since it remains weight-1 on each code block under the action of the \textsc{cnot}.
Additionally, any weight-2 $X_i Z_j$ over both code blocks is correctable, even though it can spread via the transversal \cnot to a higher weight term, because the error remains at most a single $X$ and single $Z$ on each respective code block.
All errors that are stabilizer-equivalent to these are correctable; all other errors are uncorrectable.
Whenever an orbit contains mixed-weight terms, if any term in the orbit is uncorrectable then we classify the error as uncorrectable, to give the worst-case error.
The ability to distinguish fine-grained errors, such as an uncorrectable $X_0X_1$ from a correctable $X_0X_9$, is available to us through CER and is what ultimately allows us to obtain logical estimates with multiplicative precision.

We validate the evaluated total error rate of Fig.~\ref{Fig:Logical_Errors} by using the same data to perform cycle benchmarking~\cite{erhard2019characterizing}.
In this analysis, 100 distinct initial states over 7-\cnots are prepared to estimate the 2-\cnot marginals.
From this data 20 eigenvalues are randomly sampled to estimate the total gate fidelity. Resampling several times gives an estimate for the variance.  Note that cycle benchmarking usually assumes the eigenvalues are uniformly sampled at random, however, the data we are sampling over is only a subset of the possible 14-qubit Pauli operators.
Nonetheless, this method provides an estimate for the infidelity that can be compared with our GRF model.
Using this method estimates a total error of $0.362 \pm 0.052$ (2 standard deviations), which is in agreement with the estimated value of our GRF model.


\section{Discussion \& outlook}
\label{sec:Outlook}

In this work we have used cycle error reconstruction to characterize the error profiles of single and transversal \cnots between two code blocks of the 7-qubit Steane code in a register of 16~ions.
In our characterizations, we found that the error profiles of individual \cnots can vary depending on their specific support on the register, which is in line with expectations based on device physics.
Consequently, this implies that attempts to predict logical performance from physical components should be done with explicit consideration of the context, such as how the component gates vary across the register.
Characterizing the noise profile of all constituent components of a logical operation with characterization tools like CER can be important for predicting logical performance.

Regarding the noise profile of our transversal \textsc{cnot}s, our results obtained with CER validate our model that predicts system-level performance using single component-level characterization together with a simple device physics model of time-dependent dephasing.
Simultaneously, we have demonstrated diagnostic utility by identifying and removing several experimental error sources using the data provided by CER.
Using a physically motivated correlation model together with a Gibbs random field approach, we are able to make further predictions about expected logical level performance based on physical error characterization up to 2-\cnot error terms only.

CER uses twirling to guarantee that the average-error channel studied has the required stochastic Pauli form. Coherent errors are averaged out as part of this process; while this leads to considerable simplification and increased efficiency in our error characterization, it can complicate the identification of certain physical error sources. Promisingly, as demonstrated here, identifying coherence in error models is still possible with CER, despite not using tailored approaches for estimating coherent and incoherent contributions to the error profile such as in Ref.~\cite{carignan2024estimating}.
Our demonstrated ability to identify coherent terms without additional sequences suggests that there may be ways to further reduce the experimental resources for assessing coherence, or highlight in advance the likely candidates for coherent noise that additional experimental sequences can target.

The characterizations of this work include 1-qubit, 2-qubit and 4-qubit error marginals obtained over various experimental configurations.
Ultimately, these characterizations are made possible because the amount of information obtained via CER is adjustable to the amount of information desired by the experimentalist, requiring less time resources when obtaining less information.
Using the reasonable assumption that high-weight errors are less likely than low-weight errors, marginalization captures the pertinent features of the full error profile.
This is a significant advantage over other QCVV methods, where either the full process matrix or an average error is extracted.
For large system sizes in particular, characterizing the full process matrix is usually unfeasible.
CER remains effective at learning low-weight errors at large system sizes, because disjoint marginal terms can be extracted in parallel.

The time budget required to run CER on our trapped-ion hardware is on the order of hours. Any prospect of performing CER as part of the regular diagnostics of our device would require a drastic reduction in acquisition time, beyond what can be achieved by the resource optimization performed here.
A major driver in the overall time budget is classical processing and communications delay in the experimental control electronics and software. In the superconducting community there exist hardware implementations where running parametrized circuits~\cite{rajagopala2024hardware} and randomized compiling~\cite{fruitwala2024hardware} add no overhead compared to running deterministic sequences.
With the availability of such hardware, the overall time savings for our system would be orders of magnitude and the ideal experimental parameters in shots per randomization and total randomizations would change drastically. Wall-clock time is related to the available experimental repetition rate and so will depend on the hardware platform and implementation at large. Cutting down on experimental time per shot also allows for the allocation of more shots in total, which opens up the possibility to study crosstalk over more qubits by considering larger marginals or to sample 2-\cnot error terms at higher accuracy.

We have demonstrated a physical-level characterization and used this data for predictions of logical performance. It is also possible to perform CER directly at the logical level. The studied hard cycles would then be logical gadgets, which may include mid-circuit measurements. There are now a number of different benchmarking routines for mid-circuit measurements~\cite{govia2023randomized,zhang2024generalized,Hines2025,harper2025characterising} that could enable logical characterization in tandem with ongoing error correction. Such methods focus on logical noise at the expense of explicit physical-level characterizations, which has been a core focus of this work in addition to the specific application of estimating logical error rates.
One challenge for physical-level characterizations as system sizes scale up is that the weight of relevant error terms increases in tandem with the system size.
By working with logical code states, and taking marginalizations specifically tailored for the chosen code, it is probable that GRF models could be employed in a similar fashion, using physical-level characterizations to make logical-level predictions in larger codes scalably.


\acknowledgments
The authors would like to thank Arnaud Carignan-Dugas for helpful discussions regarding cycle error reconstruction and Evan Hockings for fruitful discussions on estimating marginal probabilities. The authors are thankful to Markus M\"uller for fruitful initial discussions and the Institute for Quantum Information at RWTH Aachen for their hospitality during a research visit by N.F. where this project was initiated.

We gratefully acknowledge support by the European Union’s Horizon Europe research and innovation program under Grant Agreement Number 101114305 (“MILLENION-SGA1” EU Project), the US Army Research Office through Grant Number W911NF-21-1-0007, the Austrian Science Fund (FWF Grant-DOI 10.55776/F71) (SFB BeyondC), the Office of the Director of National Intelligence (ODNI), Intelligence Advanced Research Projects Activity (IARPA), under the Entangled Logical Qubits program through Cooperative Agreement Numbers W911NF-23-2-0216 and W911NF-23-2-0223. The Innsbruck team further received support from the IQI GmbH. M.R. acknowledges support from the Deutsche Forschungsgemeinschaft (DFG, German Research Foundation) under Germany’s Excellence Strategy Cluster of Excellence Matter and Light for Quantum Computing (ML4Q) EXC 2004/1 390534769.This research is also part of the Munich Quantum Valley (K-8), which is supported by the Bavarian state government with funds from the Hightech Agenda Bayern Plus.

The views and conclusions contained in this document are those of the authors and should not be interpreted as representing the official policies, either expressed or implied, of IARPA, the Army Research Office, or the U.S. Government. The U.S. Government is authorized to reproduce and distribute reprints for Government purposes notwithstanding any copyright notation herein. Views and opinions expressed are those of the author(s) only and do not necessarily reflect those of the European Union or the European Commission. Neither the European Union nor the granting authority can be held responsible for them.

\section*{Conflicts of interest}
T.M. is CEO at Alpine Quantum Technologies GmbH, a commercially oriented quantum computing company.

\section*{Author contributions}
N.F. and R.F. are joint lead authors. N.F. led and motivated the theoretical aspects of the work, R.F. led and conducted the experimental aspects of the work.
D.S.~contributed to the analysis of the experiment, validation of the error model assumptions, and estimation of the correctable vs uncorrectable errors.
A.S.~contributed to building and maintaining the experimental apparatus.
Ch.D.M.~supervised the experimental team, and lead the experimental team in writing and review of the manuscript.
M.R.~motivated and initiated the project and contributed to theoretical methods and data analysis. R.H.~contributed to the theoretical approach and methods.
T.M., J.E., and S.B. led the conceptualization of the research and coordinated research activities across the institutions.  
All the authors contributed to the analysis and interpretation of results, and the writing of the manuscript.

\section*{Data Availability Statement}
Data and analysis scripts supporting the findings of this study are openly available~\cite{dataset}.

\newpage
\newpage

%


\newpage

\appendix

\begin{figure*}[ht!]
\centering
\includegraphics[width = 17cm]{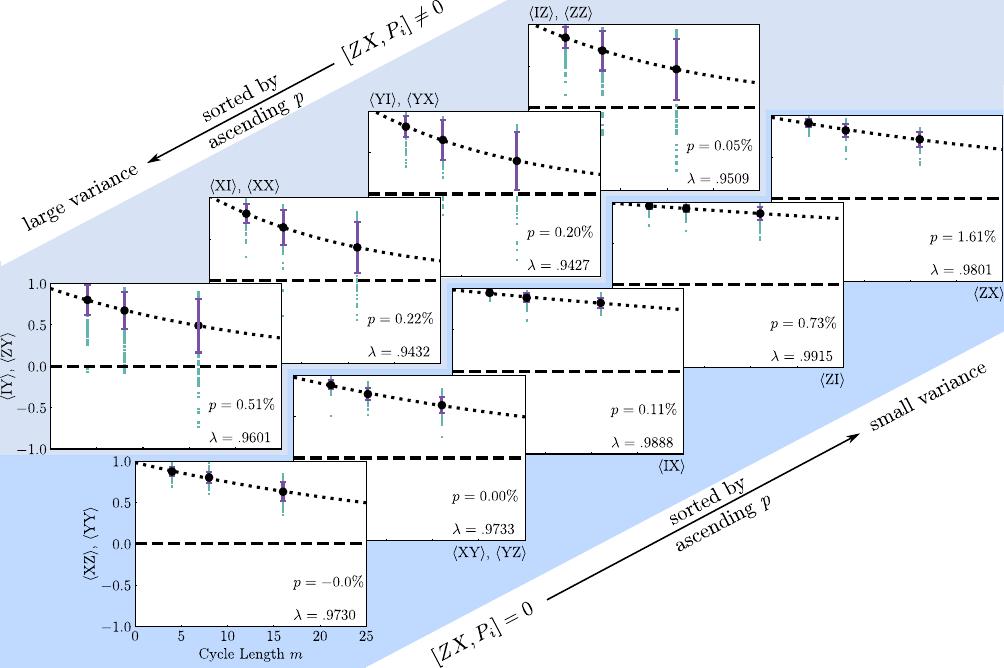}
\caption{Unique expectation value decay curves for single \cnot experiment showing signature of coherent $ZX$ miscalibration. The expected value of Pauli operators that anticommute (graphs with light shading) with $ZX$ show large variance, whereas those that commute with $ZX$ (graphs with dark shading) show small variance. All graphs share axes ranges, mean (black disc) and standard deviation (purple bars) of the expectation values of randomized circuits are indicated per sequence length $m$, and graphs are sorted by the calculated marginal probabilities; not the decay parameters. Data displayed is for \cnot with support $(0,9)$ for 40 randomizations per initial state and 150 shots per randomization, with $p_{II}=96.54\%$.}
\label{fig:ZX_Decays}
\end{figure*}

\section{Identifying the origin of $ZX$ errors}
\label{sec:ZX_Errors}

The averaged error channel obtained after twirling contains no coherent elements.
An important question for diagnostics is therefore how we can discern coherent errors using CER data.
In Sec.~\ref{sec:Single_CNOT} we presented single \cnot data that showed prominent $ZX$ errors, which we ultimately traced to a miscalibration.
These miscalibrations are coherent errors that are not discernible in the eigenvalue decay alone.
However, the error channel in each randomized instance of an experiment still has coherent terms and only in the average-error channel do we obtain stochastic behavior.
This means that when looking at individual data points rather than just the average we can glean more information.
Of course, individual inspection is not a particularly scalable procedure, where we would refer to novel work to do so more efficiently~\cite{carignan2024estimating}.

Nonetheless, it is instructive to show how we can still gain this information from the existing data, shown in Fig.~\ref{fig:ZX_Decays}.
The decay curves of Pauli expectation values are used to estimate the eigenvalues.
In this case, we can see that all the unique decay curves fall in one of two groups: Those that commute with $ZX$ show slow decay and small scatter in the individual data points around the mean.
All Pauli operators that do not commute with $ZX$ show large variation around the mean value and have relatively fast decays.
Factors that contribute to the variance of expectation values include the shot noise of individual experimental instances as well as the underlying true expectation value of each instance.
The spread of data in Fig.~\ref{fig:ZX_Decays} extends well into the negative expectation values for comparatively low shot noise, which is associated with the build up of coherent noise and would not occur with incoherent noise.
This is a typical signature of coherent errors in twirled channels, as was discussed in the literature in the context of randomized benchmarking~\cite{ball2016effect,edmunds2020dynamically}.
When coherent noise commutes with a particular Pauli it leaves the corresponding Pauli eigenstate invariant up to global phase, so these coherent noise signatures are not present in the relevant decays.

\end{document}